\documentclass[10pt, aps, prb, twocolumn, showpacs, floatfix, superscriptaddress]{revtex4-1}

\usepackage{amsmath, amssymb, mathtools, array, color, bbm}
\usepackage{hyperref}
\usepackage{graphicx}
\usepackage{times}

\newcommand{ \ex }				[1] { \langle{#1}\rangle }

\newcommand{ \ac }				[1] { \left\{{#1}\right\} }
\newcommand{ \com }			[1] { \left[ {#1} \right] }

\newcommand{ \op }				{ {p} }
\newcommand{ \ox }				{ {x} }

\newcommand{\rr}				{\mathbbm{R}}

\newcommand{\id}				{\mathbbm{1}}

\newcommand{\ve}				{\varepsilon}

\mathchardef\minus="002D

\newcommand{\uu}				{\uparrow}
\newcommand{\dd}				{\downarrow}
\newcommand{\pd}				{\phantom{\dag}}

\newcommand{\trm}				{\textrm}

\newcommand{ \tr }				{{\rm{Tr}}}

\newcommand{\nn}				{\nonumber}

\newcommand{\fc}				{\Gamma}

\newcommand{\fk}				{\mathcal{K}}

\begin{document}


\title{Entanglement of nanoelectromechanical oscillators by Cooper-pair tunneling}

\author{Stefan~Walter}
\affiliation{Institute for Theoretical Physics and Astrophysics, University of W{\"u}rzburg, 97074 W{\"u}rzburg, Germany}
\affiliation{Department of Physics, University of Basel, Klingelbergstrasse 82, CH-4056 Basel, Switzerland}
\author{Jan~Carl~Budich}
\affiliation{Department of Physics, Stockholm University, Se-106 91 Stockholm, Sweden}
\author{Jens~Eisert}
\affiliation{Qmio Group, Dahlem Center for Complex Quantum Systems, Freie Universit{\"a}t Berlin, 14195 Berlin, Germany}
\author{Bj{\"o}rn~Trauzettel}
\affiliation{Institute for Theoretical Physics and Astrophysics, University of W{\"u}rzburg, 97074 W{\"u}rzburg, Germany}

\date{\today}


\begin{abstract}
We demonstrate that entanglement of two macroscopic nanoelectromechanical resonators -- coupled to each other via a
common detector, a tunnel junction -- can be generated by running a current through the device. We introduce a setup that
overcomes generic limitations of proposals suggesting to entangle systems via a shared bath. At the heart of the proposal
is an Andreev entangler setup, representing an experimentally feasible way of entangling two nanomechanical
oscillators. Instead of relying on the coherence of a (fermionic) bath, in the Andreev entangler setup, a split Cooper-pair that
coherently tunnels to each oscillator mediates their coupling and thereby induces entanglement between them. Since
entanglement is in each instance generated by Markovian and non-Markovian noisy open system dynamics in an
out-of-equilibrium situation, we argue that the present scheme also opens up perspectives to observe dissipation-driven
entanglement in a condensed-matter system.
\end{abstract}

\maketitle

\section{Introduction}
\label{sec:one}

One of the most fascinating perspectives offered by nanomechanical devices is the creation of entanglement in systems exhibiting
macroscopic length scales. The quest for observing such genuine quantum effects in macroscopic physical systems is not only
motivated by fundamental considerations, relating to long-standing questions of the quantum-to-classical transition. But in fact,
several applications for quantum technologies are readily conceivable: Mechanical systems may be used in quantum metrology
and high-precision measurement, and possibly even in quantum interfaces for architectures of quantum information processing.
Indeed, only within the last few years, the experimental study of nanoelectromechanical \cite{Electromechanics1,Electromechanics2,Electromechanics3,Electromechanics4} and optomechanical
\cite{Optomechanics1,Optomechanics2,Optomechanics3,Optomechanics4} systems close to the quantum regime has seen significant successes, one of the latest experimental achievements
constituting the cooling of a nanomechanical resonator to close to its ground state \cite{GroundStateCooling1,Electromechanics1,GroundStateCooling3}. Still, the benchmark of
achieving entanglement, from which a number of exciting developments would follow, has not quite been reached yet, neither in
the nanoelectromechanical nor in the optomechanical setting.

On the theoretical side, to devise feasible proposals for entanglement generation in nanoelectromechanical systems (NEMS)
has been a goal for many years. In the pioneering work of Ref.~\onlinecite{Eisert:2004aa}, a route towards entanglement in NEMS
was proposed making use of a global nonadiabatic change of the interaction strength in a one-dimensional chain of
nanomechanical oscillators. Most recent proposals focus rather on optomechanical settings. Here, the generation of
entanglement between a movable mirror and a cavity \cite{Vitali:2007is,Cavities} or with another mirror \cite{MirrorMirrorEntanglement1,MirrorMirrorEntanglement2,MirrorMirrorEntanglement3}
(mediated by cavity modes) has been studied. Other proposals consider entanglement with a collective spin ensemble of
an atomic medium \cite{Hammerer:2009tf} or a Bose-Einstein  condensate \cite{BEC1,BEC2}. The aim is in any of these proposals
to detect the generation of macroscopic entanglement (as is already routinely experimentally observed in quantum optical
systems of atoms and photons or collective spin ensembles); still, for technological reasons, it remains a challenging task.

\begin{figure*}[ht]
	\center{\includegraphics[width=0.8\textwidth]{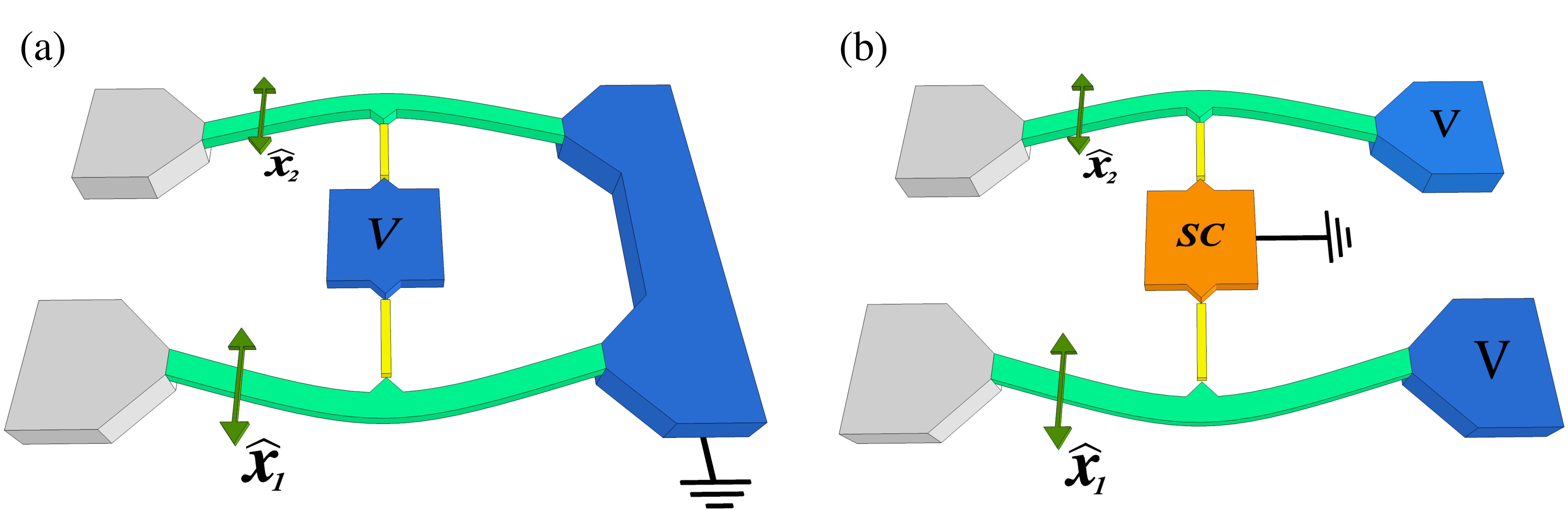}}
	\caption{\label{fig:2oscSetups}(Color online)
	(a) Schematic setup of a bipartite continuous variable quantum system, realized as two nanoelectromechanical oscillators
	(green) in a tunnel junction setup which share common fermionic reservoirs (blue). Yellow lines indicate tunnel junctions.
	(b) Two nanomechanical oscillators are effectively coupled in an Andreev entangler setup due to a Cooper-pair that is split.
	Each of the two electrons of the Cooper-pair is assumed to tunnel from the superconductor (orange) onto a different oscillator.
	}
\end{figure*}

Our aim is to generate entanglement between two spatially separated nanomechanical oscillators via open systems dynamics
moderated by a shared bath. On an abstract level, several such settings have been studied so far, mainly considering couplings
to bosonic baths \cite{CommonBath1,CommonBath2}. To uplift such suggestions to feasible schemes, however, one has to acknowledge the
obstacle that in order to achieve entanglement, bath coherence times have to be long. Additionally, even more crucial challenges
have to be overcome: In spatially separated tunneling processes, only very little which-path information may be acquired, if
entanglement is not to be largely destructed. This is a generic fundamental issue in all systems separated on macroscopic
length scales and coupled by true multimode quantum baths (which is in sharp contrast with optical setups, where, say, several
atoms can easily couple to the same optical field mode of a cavity). We solve this problem by suggesting an entirely new mechanism for entanglement generation between nanoelectromechanical oscillators: We resort to an Andreev entangler setup which is characterized by the manifest
absence of which-path information in a certain parameter regime, see Fig.~\ref{fig:2oscSetups}(b). This is reached by using a dynamical
instance of a Coulomb blockage and employing the protection against uncorrelated quasiparticle tunneling offered by the superconducting gap. For a split Cooper pair,
the electrons are found to tunnel separately into different leads. Due to electron-phonon coupling, this process of Cooper-pair tunneling effectively generates robust entanglement between the bosonic mechanical modes. Hence, our
proposal only relies on the coherence of a superconducting condensate, exhibiting much longer coherence lengths
than normal metal systems. We quantify this entanglement as it evolves in time under Markovian \cite{Markovian} and
non-Markovian \cite{NonMarkov1,NonMarkov1b,NonMarkov2,NonMarkov3,NonMarkov4} dynamics. It has increasingly become clear in recent years that dissipation is not
necessarily always detrimental for the generation of entanglement or coherent dynamics \cite{Dissipation1,Dissipation2,Dissipation3}. For such
an idea to work for tunneling processes in the condensed-matter context, the above mentioned challenges have to be
overcome, however. In this work, we describe a way in which this could be done.

In Sec.~\ref{sec:two}, we introduce the two settings under consideration in this article on a general footing. We continue
in Sec.~\ref{sec:three} with a detailed study of our first setup, setup (A), which is capable of generating entanglement
between two nanomechanical oscillators which share a common fermionic reservoir. Section~\ref{sec:four} is devoted to
the heart of our article, setup (B), which introduces a new mechanism, relying on the physics of an Andreev entangler
to generate entanglement between the two nanomechanical oscillators via correlated fermion tunneling. Finally, we conclude in Sec.~\ref{sec:end} and present details on the entanglement dynamics in the appendices.

\section{Basic setting}
\label{sec:two}

Both of our proposed setups are captured by generic Hamiltonians of the form 
\begin{equation}
	H = H_{S} + H_{B} + H_{SB}, 
\end{equation}
where $H_{S}$ and $H_{B}$ describe
the system and bath degrees of freedom, respectively. The system Hamiltonian is given by 
\begin{equation}
	H_{S} = \sum_{i=1,2} \op_{i}^{2}/2 m_{i} +m_{i} \Omega_{i}^{2} \ox_{i}^{2}/2
\end{equation}	
which corresponds to two nanomechanical bosonic oscillators with effective mass $m_{i}$, frequency $\Omega_{i}>0$, and
position and momentum operators $\ox_{i}$ and $\op_{i}$, respectively. For simplicity, we later assume two identical oscillators,
i.e., $\Omega_{1}=\Omega_{2}=\Omega$, and $m_{1}=m_{2}=m$, but this assumption is not crucial for the generation of
entanglement in our setup. The coupling between the system and the bath is defined by $H_{SB}$. Figure~\ref{fig:2oscSetups}
depicts two settings, setup (A) -- which may be seen as a paradigmatic setup coupling two modes via a shared fermionic bath
-- and setup (B), which constitutes the Andreev entangler setup in the focus of this work. Our goal is to investigate the time
evolution of entanglement between two nanomechanical oscillators. For this purpose, we employ the two most used
entanglement measures, the entanglement of formation $E_F$ \cite{EOF,GEOF} as well as the (logarithmic) negativity
$E_N$ \cite{Negativity1,Negativity1b,Negativity2,Negativity3}; both are equipped  with an operational interpretation. The latter quantity is given by $E_N(\rho_S ) = \log_2(\| \rho_S^\Gamma\|_1)$, $\rho_S^\Gamma$
being the partial transpose of the state of the system, while $E_F$ is the convex hull of the reduced entropy function; see Appendix~\ref{sec:appendix}.
In all instances, the dynamics we consider preserves the Gaussian character of initial Gaussian states. For Gaussian states
with vanishing first moments, the computation of the entanglement measures is particularly simple. Here, both measures can
be computed from the symmetric $4\times4$-covariance matrix $\fc(t)$ of the state, with entries 
\begin{equation}	
	\fc_{j,k}= \tr(\rho_S (t) \{R_j, R_k \}),
\end{equation}	
where the vector of quadratures is given by ${R} = (\ox_{1},\op_{1},\ox_{2},\op_{2})^{T}$. We compute the time dependence
of the entries of $\fc(t)$ with an equation of motion for the system's density matrix $\rho_{S}(t)$, where we capture the
non-Markovian regime by employing a time convolutionless master equation method \cite{Breuer:2002wp}. The non-Markovian
dynamics implies that the system's time evolution depends on its history, reflected in the equation of motion by time-dependent
damping and decoherence kernels. The master equation in the Born approximation reads (we put $\hbar=1$)
\begin{align}\label{eqn:nemsEnt4}
	\dot{\rho}_{S}(t) =	&-i \com{H_{S},\rho_{S}(t)}  \\
					- \int_{0}^{t} d\tau \, & \tr_{B} \left[ H_{SB}, \left[ H_{SB}(\tau-t),\rho_{S}(t) \otimes \rho_{B} \right]  \right] \, . \nn
\end{align}
After having introduced the generic Hamiltonian and the measures of entanglement which we use, we continue with a
detailed study of setup (A).

\section{Coupling via shared fermionic baths}
\label{sec:three}
We start by discussing the general mechanism of entanglement generation via shared fermionic baths, referred to as setup (A).
Even if it suffers from the above-mentioned drawbacks, it still exemplifies some of the basic principles, and serves as the
paradigmatic setting on which our proposal, setup (B), the Andreev entangler setup, is built. setup (A) consists of two nanoelectromechanical
oscillators that are both coupled via an atomic point contact to common electronic reservoirs; cf. Fig.~\ref{fig:2oscSetups}(a).
Importantly, the oscillators are not directly coupled to each other (as, e.g., in Ref.~\onlinecite{Ludwig:2010wn}), but only through the
electron reservoirs which act as fermionic baths. The yellow tunnel junctions in Fig.~\ref{fig:2oscSetups}(a) are sufficient to
generate an entangled state of the two oscillators. Here, 
\begin{equation}
	H_{B} = \sum_{r} \ve_{r}^{\pd} \psi^{\dag}_{r} \psi^{\pd}_{r} + \sum_{l} \ve_{l}^{\pd} \psi^{\dag}_{l} \psi^{\pd}_{l}
\end{equation}	
refers to the two leads, depicted blue in Fig.~\ref{fig:2oscSetups}(a). $\psi_{i}^{(\dag)}$ are electron annihilation (creation)
operators and $i=l(r)$ is a wave vector in the left (right) reservoir. (Note that the left reservoir denotes the central island and the right reservoir the large electrode connecting the right-hand sides of the two oscillators.) The system-bath coupling is meditated via the tunneling
Hamiltonian
\begin{align}\label{eqn:nemsEnt8}
	H_{SB} = \sum_{l,r} \sum_{n=1,2} (\gamma_{0}+\gamma_{x} {x}_{n}) \psi_{l}^{\dag} \psi^{\pd}_{r} + \trm{H.c.} \, ,
\end{align}
where for small oscillation amplitudes we approximate the tunneling matrix elements as being linear in the oscillator displacement,
and for the sake of simplicity assume all tunneling amplitudes ($\gamma_{0}$ and $\gamma_{x}$) to be real and take the coupling
to be symmetric.
As one can see from Eq.~(\ref{eqn:nemsEnt8}), we also neglect the momentum dependence of
the tunneling amplitudes. Therefore, which-path information is effectively
discarded at the Hamiltonian level, meaning that after the tunneling from
the left reservoir to the right one, there is no information included in
the model on the oscillator that was involved in the tunneling process.
For a setup of finite size -- like the one shown in Fig.~\ref{fig:2oscSetups}a -- this is a
rather strong approximation since the geometry, especially the fact that
the right reservoir is of mesoscopic size, is not taken into account at
all. This strong assumption is however weakened when we introduce setup (B)
below.
In order to include non-Markovian effects of the fermionic reservoirs properly, we need to include an energy-dependent density of states in the leads. This is done with a Lorentzian-shaped density of states which means that an electron with
energy $\ve_{l}$ in the left lead can tunnel into states of the right lead with energy $\ve_{r}$, broadened by $L_{c}$ \cite{Liu:2007aa,WingreenEtAl1,WingreenEtAl2,WingreenEtAl3}.
The equation of motion for the state of the system up to second order in the tunneling can then be written as
\begin{align}\label{eqn:nemsEnt10}
	\dot{\rho}_{S}(t) = &	- i \com{H_{S} + i \fk_{-}^{(2)}(t) ({x}_{1} + {x}_{2})^{2} , \rho_{S}(t)} \nn \\
				&	- \fk_{+}^{(1)}(t) \com{{x}_{1} + {x}_{2}, \com{{x}_{1} + {x}_{2}, \rho_{S}(t)} } \nn \\
				&	+ \tilde{\fk}_{+}^{(1)}(t) \com{{x}_{1} + {x}_{2}, \com{{p}_{1} + {p}_{2}, \rho_{S}(t)} } \nn \\
				&	+ \tilde{\fk}_{-}^{(2)}(t) \com{{x}_{1} + {x}_{2}, \ac{{p}_{1} + {p}_{2}, \rho_{S}(t)} } \, ,
\end{align}
where the damping and decoherence kernels are given in the next section.

\subsection{Damping and decoherence kernels of setup (A)}
\label{sec:threeOne}

We can deduce the appearing kernels for setup (A) in Eq.~(\ref{eqn:nemsEnt10}) by first simplifying the system-bath coupling
Hamiltonian and write
\begin{align}\label{eqn:supm1}
	H_{SB} = \sum_{n=1,2} S_{n} E + S_{n}^{\dag} E^{\dag} \, ,
\end{align}
where $S_{n}$ and $E$ denote arbitrary operators acting only on the Hilbert space of system and bath, respectively.
We define the system and bath operators as $S_{n} = \gamma_{0}+\gamma_{x} {x}_{n}$ and $E=\psi_{l}^{\dag} \psi^{\pd}_{r}$,
respectively. With this, the equation of motion for the reduced density matrix of the system up to second order in the tunneling term can be written in the form of Eq.~(3), where we 
have defined the time-dependent damping and decoherence kernels
\begin{align}
	\fk_{+}^{(1)}(t) &= \int_{0}^{t} d\tau \left( K^{(1)}(\tau) + K^{(1)}(-\tau) \right) \gamma_{x}^{2} \cos(\Omega \tau) 						\, , \\
	\tilde{\fk}_{+}^{(1)}(t) &= \int_{0}^{t} d\tau \left( K^{(1)}(\tau) + K^{(1)}(-\tau) \right) \frac{\gamma_{x}^{2}}{m \Omega} \sin(\Omega \tau) 	\, , \\
	\fk_{-}^{(2)}(t) &= \int_{0}^{t} d\tau \left( K^{(2)}(\tau) - K^{(2)}(-\tau) \right) \gamma_{x}^{2} \cos(\Omega \tau) 						\, , \\
	\tilde{\fk}_{-}^{(2)}(t) &= \int_{0}^{t} d\tau \left( K^{(2)}(\tau) - K^{(2)}(-\tau) \right) \frac{\gamma_{x}^{2}}{m \Omega} \sin(\Omega \tau)		
\end{align}
with
\begin{align}
	K^{(1)}(t) &= \frac{1}{2} \ex{\ac{E(t),E^{\dag}(0)}}		\label{eqn:supm3a} \, ,\\
	K^{(2)}(t) &= \frac{1}{2} \ex{\com{E(t),E^{\dag}(0)}} 		\label{eqn:supm3b} \, .
\end{align}
We see from the equation of motion for $\rho_{S}(t)$ that its time evolution is governed by the time-dependent damping
and diffusion kernels $\fk_{+/-}^{(1/2)}(t)$ and $\tilde{\fk}_{+/-}^{(1/2)}(t)$. We will briefly sketch the calculation of the kernels
$K^{(1/2)}(t)$. Due to the fact that the equation of motion for $\rho_{S}(t)$ is of second order in $H_{SB}$, the only possibility
of including non-Markovian effects is by considering an energy-dependent density of states in the leads. Including
non-Markovian effects leads to a finite correlation time in the leads, and is a key ingredient for the entanglement. We mention here that
non-Markovian effects in the reservoirs and their influence on entanglement of two quantum systems have previously been studied
in Refs.~\onlinecite{Liu:2007aa,YingHua:2010uf,Xiao:2010gm,Bellomo:2010cn}. The kernels
$K^{(m)}(t)$
are given by
\begin{align}
	K^{(m)}(t) 	&= \frac{1}{2} \int d\ve_{l} \int d\ve_{r} \, J(\ve_{l},\ve_{r}) \, e^{i (\ve_{l}-\ve_{r})t} \\
		&\big[ n(\ve_{l}) (1-n(\ve_{r})) - (-1)^{m} n(\ve_{r}) (1-n(\ve_{l})) \big] \, ,\nonumber
\end{align}
where $n(\ve_{x}) = (e^{\beta(\ve_{x}-\mu_{x})}+1)^{-1}$ is the Fermi distribution function and we introduce an energy-dependent
spectral function
\begin{align}
	J(\ve_{l},\ve_{r}) = \sum_{k,q} \delta(\ve_{l}-\ve_{k}) \delta(\ve_{r}-\ve_{q}) \, .
\end{align}
To account for a finite lifetime of quasiparticles in the leads, the $\delta$-functions are smeared out and replaced by
Lorentzians of width $L_{c}$:
\begin{align}\label{eqn:supm6}
	J(\ve_{l},\ve_{r}) =\sum_{k,q} \frac{L_{c}}{(\ve_{l}-\ve_{k})^{2}+L_{c}^{2}} \frac{L_{c}}{(\ve_{r}-\ve_{q})^{2}+L_{c}^{2}} \, .
\end{align}
The largest contribution of each of the independent sums in Eq.~(\ref{eqn:supm6}) will come from the energies close to the Fermi
level of each lead. We further restrict ourselves to the regime of low applied bias voltages ($V<L_{c}$). This is just a simplification
in order to keep the number of parameters in the problem as low as possible. With these assumptions, the energy-dependent
spectral function can be approximated as
\begin{align}\label{eqn:supm7}
	J(\ve_{l},\ve_{r}) = \frac{1}{(\ve_{l}-\ve_{r})^{2}+L_{c}^{2}} \, .
\end{align}
Physically, Eq.~(\ref{eqn:supm7}) implies that an electron with energy $\ve_{l}$ in the left lead can tunnel into states of
the right lead with energy $\ve_{r}$, broadened by $L_{c}$ \cite{WingreenEtAl1,WingreenEtAl2,Liu:2007aa,WingreenEtAl3}. Let us briefly interpret the physical meaning of the parameter $L_{c}$. On the one hand, the limit $L_{c} \rightarrow 0$ corresponds to a resonant tunneling process with narrow densities of states in the leads. On the other hand, the so-called wide-band limit is reached in the limit $L_{c} \rightarrow \infty$. This limit results in an energy-independent density of states in the leads. Thus, in that latter case, any electron from the left
lead can tunnel into the right lead. In nature, a realistic form of the density of states depends on the details and can in many situations not be described by our model with a single parameter $L_c$. However, this model nicely allows us to extrapolate between two extreme cases. Therefore, we can better understand qualitative aspects of the role of the density of states in the leads for the generation of entanglement.

With the spectral function in Eq.~(\ref{eqn:supm7}), the time-dependent kernels $\fk_{+/-}^{1/2}(t)$ and $\tilde{\fk}_{+/-}^{1/2}(t)$
can now be calculated analytically, but the resulting expressions are too lengthy to be stated here.

\subsection{Non-Markovian entanglement dynamics for setup (A)}
\label{sec:threeTwo}

\begin{figure}[ht]
\centering{
	\includegraphics[width=0.95\columnwidth]{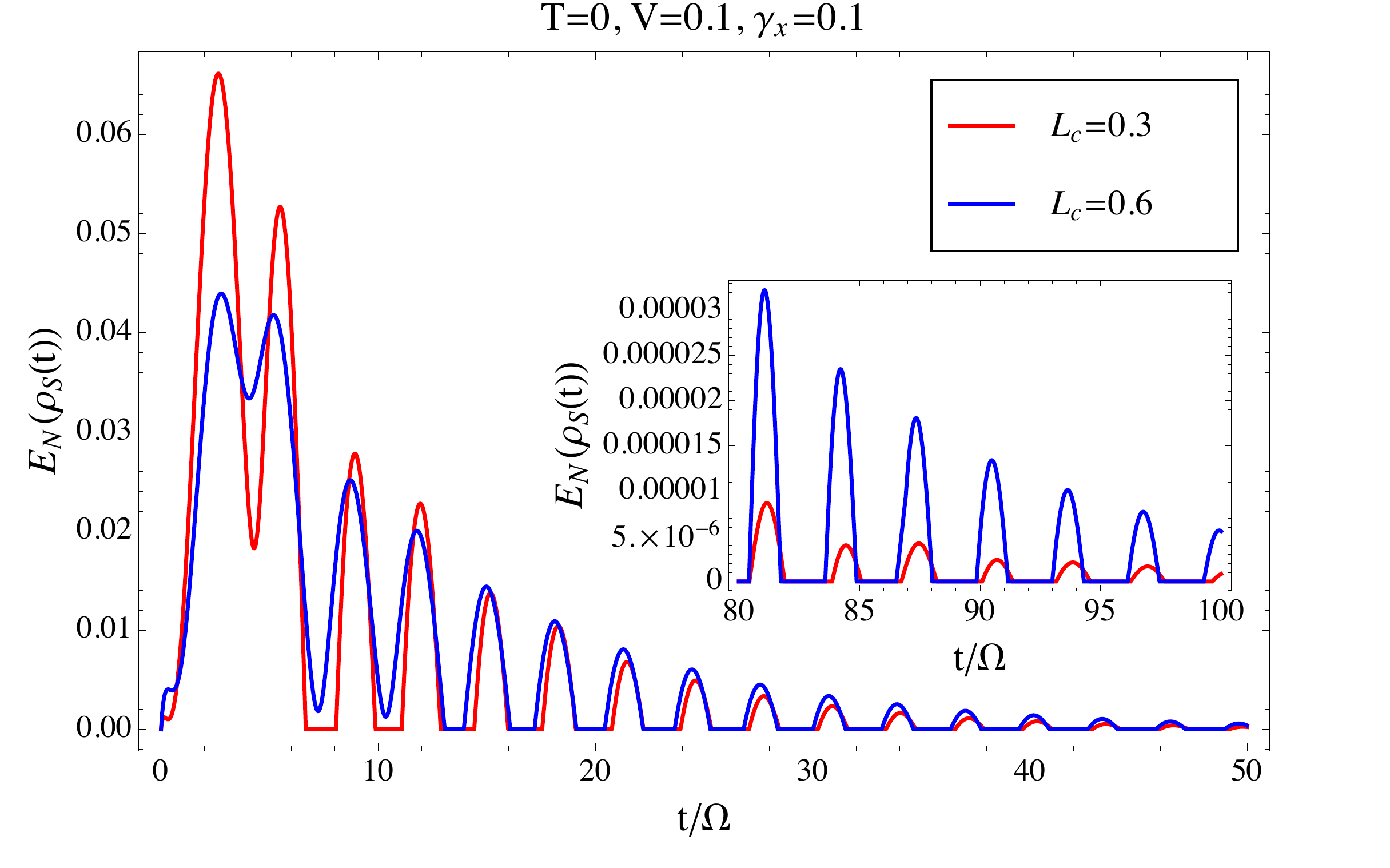}
	\caption{(Color online) Logarithmic negativity for $L_{c}=0.3 \, \Omega$ (red) and $L_{c}=0.6 \, \Omega$ (blue). Here, the
	temperature is $T=0$,
	$V=0.1 \, \Omega$, and $\gamma_{x}=0.1 \sqrt{m\Omega}$. As can be seen, the qualitative dependence on the parameter $L_{c}$ is rather small. }
	\label{fig:comF3}
	}
\end{figure}

We now investigate the system's dynamics, where we take, for simplicity, the idealized vacuum state as an initial state, which is a
Gaussian state. We first discuss the case of zero electronic temperature. In Fig.~\ref{fig:comF3} we show the logarithmic negativity
for $\gamma_{x} = 0.1  \sqrt{m\Omega}$ and its dependence on $L_{c}$ for setup (A). We clearly see that the initially separable
state becomes entangled right after the interactions have been suddenly switched on. This effect of the sudden quench leading
to a nonequilibrium situation is similar to the nonadiabatic change of the interaction strength in Ref.~\onlinecite{Eisert:2004aa},
but here mediated via a shared bath. Experimentally, the switching in this tunneling setup can be achieved by gates controlling
the tunnel coupling via a resonance. Physically this means that the rise time of the gates has to be shorter than $1/\Omega$, the
timescale of the oscillators. Typical rise times of electronic gates can be as short as $60\, \trm{ps}$ \cite{Dovzhenko:2011eg}.
For an oscillator with a resonance frequency of $\sim 500 \, \trm{MHz}$ (which can in principle be brought into its ground state
at low temperatures) these rise times are sufficient to accomplish the sudden switching. In its subsequent evolution, the
entanglement oscillates in time, before it slowly decays. We further find that low bias voltages show an increased logarithmic
negativity compared to high bias voltages.
%

\section{The Andreev entangler setup}
\label{sec:four}

A significant challenge in the idealized setup (A) is the assumption of long coherence times of the reservoirs.
To circumvent this problem, we now introduce an alternative realization based on a superconducting reservoir in the center
and two independent normal-metal reservoirs which nonetheless is capable of generating entanglement. This realization,
setup (B), is of an Andreev entangler type.

In its original sense, an Andreev entangler \cite{Recher:2001je} relies on the splitting
of a Cooper pair which is a spin singlet. There, the split Cooper pair stemming from an $s$-wave superconductor can coherently tunnel
via two different quantum dots into different leads. Throughout this process, the spin singlet is preserved and therefore the
Andreev entangler is capable of generating nonlocal spin-entangled electrons.

In our context of NEMS, we use only charge
properties of the split Cooper pair as a mediator for an effective coupling between the oscillators. We show in Fig.~\ref{fig:2oscSetups}(b)
what such an Andreev entangler setup could look like. A superconducting island (orange) serves as a source of Cooper pairs
which can tunnel onto two different (conducting) nanomechanical oscillators. The process where the Cooper pair is split and
one electron tunnels to the lower oscillator and the other one tunnels to the upper oscillator gives rise to an effective coupling
between them.

The major improvements of setup (B) over setup (A) are
(i) the right common reservoir of setup (A) is now cut into two halves; the resulting two normal-metal leads are entirely \emph{independent} (i.e., we circumvent the assumption of macroscopic coherence in the metallic reservoirs);
(ii) both normal-metal leads see the same tunneling processes (therefore their independence does not result in which-path information) \cite{setup_a}; 
(iii) phase coherence lengths of superconductors are typically much longer than of normal metals [this allows for a larger (than pointlike) spatial extent of the center reservoir].
Processes where one Cooper pair tunnels to the upper/lower lead are not taken into account since they do not lead to any
entanglement, but rather to a background tunneling current. Such processes are also energetically not favorable due to
interaction effects in the leads and/or tunneling via a dot with a finite charging energy \cite{Recher:2001je}. Another possibility
to favor the mentioned tunneling process is to couple the superconductor to two Luttinger liquid leads \cite{Recher:2002ee}.
In NEMS this could be achieved by modeling the oscillators as one-dimensional nanowires or carbon nanotubes.

We briefly want to mention that all the ingredients needed for setup (B) have already been separately realized in recent experiments. In the last years, there have been successful experimental realizations of a Cooper-pair splitter \cite{CPS1,CPS2} where in one of the experiments carbon nanotubes were already implemented as the medium to favor the tunneling of a split Cooper pair. Suspended carbon nanotubes have also been achieved by different experimental groups worldwide. In particular, they have been studied in the context of the Franck-Condon blockade where a mechanical mode of the suspended nanotube couples to the electron occupation number of a quantum dot that forms on the carbon nanotube \cite{CNT}. Putting these two ingredients together would already allow for a first experimental test of our proposed Andreev entangler for nanoelectromechanical oscillators.

\subsection{Hamiltonian for setup (B)}
\label{sec:fourZero}

In the case of setup (B), the bath consists of three independent reservoirs, two normal-metal leads, and a central superconducting region. The
bath Hamiltonian becomes $H_{B} = H_{I} + H_{U} + H_{D}$ with 
\begin{equation}
	H_{\alpha} = \sum_{k, \sigma} \ve_{\alpha, k}^{\pd} b_{\alpha, k, \sigma}^{\dag} b_{\alpha, k, \sigma}^{\pd}
\end{equation}	
and $H_{I} = \sum_{k, \sigma} E_{k}^{\pd} \beta_{k, \sigma}^{\dag} \beta_{k, \sigma}^{\pd}$. $b_{\alpha, k, \sigma}$ are
electron annihilation operators with spin $\sigma =\uu,\dd$ and $k$ is a wave vector in lead $\alpha = U,D$. Here,
\begin{equation}
	E_{k} = ({(\varepsilon_{k} - \mu_{S})^{2} + \Delta^{2}})^{1/2}
\end{equation}	
is the quasiparticle energy and $\beta_{k, \sigma}$ is the
quasiparticle annihilation operator in the superconductor. We take the superconductor to be grounded and each lead to
be held at bias voltage $V$. The system-bath interaction is mediated by the tunneling of a split Cooper pair \cite{NoteSCgap}.
The process we focus on is therefore described by the effective tunneling Hamiltonian
\begin{align}\label{eqn:nemsEnt21}
	H_{SB} 	= \sum_{k_{1},k_{2},k_{3},k_{4}}  &T_{U}^{\pd} T_{D}^{\pd} b_{D-k_{1},\dd}^{\dag}
	b_{Uk_{2},\uu}^{\dag} c_{I-k_{3},\dd}^{\pd} c_{Ik_{4},\uu}^{\pd} \\
			+ &T_{U}^{\pd} T_{D}^{\pd} b_{Dk_{1},\uu}^{\dag} b_{U-k_{2},\dd}^{\dag} c_{I-k_{3},\dd}^{\pd} c_{Ik_{4},\uu}^{\pd}
		+ \trm{H.c.} \nn \, ,
\end{align}
where, $T_{\alpha} = \gamma_{0,\alpha} + \gamma_{x,\alpha} {x}_{\alpha}$.
The effective Hamiltonian, Eq.~(\ref{eqn:nemsEnt21}), assumes the presence of phase-coherent
Cooper pairs which break up and -- by coherent tunneling -- mediate the interaction between the two oscillators.
The quasiparticle operators of the Hamiltonian $H_{I}$ are related to electron annihilation operators through
the Bogoliubov transformation
\begin{align}
	c_{k,\uu} 	&= u_{k} \beta_{k,\uu} + v_{k}^{\pd} \beta_{-k,\dd}^{\dag}	\, , \\
	c_{-k,\dd} 	&= u_{k} \beta_{-k,\dd} - v_{k}^{\pd} \beta_{k,\uu}^{\dag} 	\, ,
\end{align}
where 
\begin{eqnarray}
	u_{k} &=& ({1/2 + \xi_{k}/(2E_{k})})^{1/2},\\
	v_{k} &=& ({1/2 - \xi_{k}/(2E_{k})})^{1/2}, 
\end{eqnarray}
with $\xi_{k} = \varepsilon_{k} - \mu_{S}$.

The effective Hamiltonian, cf. Eq.~(\ref{eqn:nemsEnt21}), exhibits two important but different contributions. The
first contribution stems from terms of order $\gamma_{0} \gamma_{x}$. These terms are qualitatively similar to terms of order
$\gamma_{x}$ in setup (A), meaning that to second order in $\gamma_{x}$ we expect a qualitatively similar behavior for the
degree of entanglement. Only considering these terms in the master equation, all kernels $\fk(t)$ and $\tilde{\fk}(t)$ for setup (B)
can be calculated in the same way as the kernels for setup (A) which we briefly sketch in the following section.

\subsection{Damping and decoherence kernels of setup (B)}
\label{sec:fourOne}

The equation of motion and the calculation of the kernels for the master equation in the case of setup (B) are formally very similar
to those of setup (A). 
As before, we introduce system and bath operators
\begin{align}
	S_{0} &= \gamma_{0U} \gamma_{0D} 					 \, , \\
	S_{1} &= \gamma_{0D} \gamma_{xU} {x}_{U} 			 \, , \\
	S_{2} &= \gamma_{0U} \gamma_{xD} {x}_{D}			 \, , \\
	S_{3} &= \gamma_{xU} \gamma_{xD} {x}_{D} {x}_{U} 	 \, ,
\end{align}
and
\begin{align}\label{eqn:supm10}
	E = b_{D-k_{1},\dd}^{\dag} b_{Uk_{2},\uu}^{\dag} c_{I-k_{3},\dd}^{\pd} c_{Ik_{4},\uu}^{\pd} \, ,
\end{align}
respectively, which allow us to write the system-bath coupling Hamiltonian in a similar way as in Eq.~(\ref{eqn:supm1}).
With this, the equation of motion for setup (B) is similar to Eq.~(3) and can be written as
\begin{align}\label{eqn:supm11}
	\dot{\rho}_{S}(t) = &-i\com{H_{S} + i \fk_{-}^{(2)}(t) ({x}_{1} + {x}_{2})^{2} , \rho_{S}(t)} \\
				&- \fk_{+}^{(1)}(t) \com{{x}_{1} + {x}_{2}, \com{{x}_{1} + {x}_{2}, \rho_{S}(t)} } \nn \\
				&+ \tilde{\fk}_{+}^{(1)}(t) \com{{x}_{1} + {x}_{2}, \com{{p}_{1} + {p}_{2}, \rho_{S}(t)} } \nn \\
				&+ \tilde{\fk}_{-}^{(2)}(t) \com{{x}_{1} + {x}_{2}, \ac{{p}_{1} + {p}_{2}, \rho_{S}(t)} } \nn \\
				&- \fk_{+}^{(3,1)}(t) \com{{x}_{1} {x}_{2}, \com{{x}_{1} {x}_{2}, \rho_{S}(t)} } \nn \\
				&- \fk_{+}^{(3,2)}(t) \com{{x}_{1} {x}_{2}, \com{{p}_{1} {p}_{2}, \rho_{S}(t)} } \nn \\
				&+ \fk_{+}^{(3,3)}(t) \com{{x}_{1} {x}_{2}, \com{{x}_{1} {p}_{2}+{x}_{2} {p}_{1}, \rho_{S}(t)} } \nn \\
				&- \fk_{-}^{(3,1)}(t) \com{{x}_{1} {x}_{2}, \ac{{x}_{1} {x}_{2}, \rho_{S}(t)} } \nn \\
				&- \fk_{-}^{(3,2)}(t) \com{{x}_{1} {x}_{2}, \ac{{p}_{1} {p}_{2}, \rho_{S}(t)} } \nn \\
				&+ \fk_{-}^{(3,3)}(t) \com{{x}_{1} {x}_{2}, \ac{{x}_{1} {p}_{2}+{x}_{2} {p}_{1}, \rho_{S}(t)} }  \nn \, ,
\end{align}
where we have defined the time-dependent memory kernels in Eq.~(\ref{eqn:supm11}) as
\begin{align}
	\fk_{+}^{(1)}(t) &= \int_{0}^{t} d\tau \left( K^{(1)}(\tau) + K^{(1)}(-\tau) \right) \gamma_{0}^{2} \gamma_{x}^{2} \cos(\Omega \tau) \, , \\
	\tilde{\fk}_{+}^{(1)}(t) &= \int_{0}^{t} d\tau \left( K^{(1)}(\tau) + K^{(1)}(-\tau) \right) \frac{\gamma_{0}^{2} \gamma_{x}^{2}}{m \Omega} \sin(\Omega \tau) \, , \\
	\fk_{-}^{(2)}(t) &= \int_{0}^{t} d\tau \left( K^{(2)}(\tau) - K^{(2)}(-\tau) \right) \gamma_{0}^{2} \gamma_{x}^{2} \cos(\Omega \tau) \, , \\
	\tilde{\fk}_{-}^{(2)}(t) &= \int_{0}^{t} d\tau \left( K^{(2)}(\tau) - K^{(2)}(-\tau) \right) \frac{\gamma_{0}^{2} \gamma_{x}^{2}}{m \Omega} \sin(\Omega \tau) \, ,
\end{align}	
as well as
\begin{align}	
	\fk_{+}^{(3,1)}(t) = \int_{0}^{t} d\tau &\left( K^{(1)}(\tau) + K^{(1)}(-\tau) \right) \\  &\times \gamma_{x}^{4} \cos(\Omega \tau) \cos(\Omega \tau) \, , \nn\\
	\fk_{+}^{(3,2)}(t) = \int_{0}^{t} d\tau &\left( K^{(1)}(\tau) + K^{(1)}(-\tau) \right) \\  &\times \frac{\gamma_{x}^{4}}{m^{2} \Omega^{2}} \sin(\Omega \tau) \sin(\Omega \tau) \, ,\nn \\
	\fk_{+}^{(3,3)}(t) = \int_{0}^{t} d\tau &\left( K^{(1)}(\tau) + K^{(1)}(-\tau) \right) \\  &\times \frac{\gamma_{x}^{4}}{m \Omega} \cos(\Omega \tau) \sin(\Omega \tau) \, \nn
\end{align}
and
\begin{align}	
	\fk_{-}^{(3,1)}(t) = \int_{0}^{t} d\tau &\left( K^{(2)}(\tau) - K^{(2)}(-\tau) \right) \\  &\times \gamma_{x}^{4} \cos(\Omega \tau) \cos(\Omega \tau) \, , \nn\\
	\fk_{-}^{(3,2)}(t) = \int_{0}^{t} d\tau &\left( K^{(2)}(\tau) - K^{(2)}(-\tau) \right) \\  &\times \frac{\gamma_{x}^{4}}{m^{2} \Omega^{2}} \sin(\Omega \tau) \sin(\Omega \tau) \, , \nn\\
	\fk_{-}^{(3,3)}(t) = \int_{0}^{t} d\tau &\left( K^{(2)}(\tau) - K^{(2)}(-\tau) \right) \\  &\times \frac{\gamma_{x}^{4}}{m \Omega} \cos(\Omega \tau) \sin(\Omega \tau) \, .\nn
\end{align}
The kernels $K^{(1/2)}(t)$ are given in Eq.~(\ref{eqn:supm3a}) and (\ref{eqn:supm3b}) with $E$ given in Eq.~(\ref{eqn:supm10}).
The actual calculation of the kernels for the Andreev entangler setup goes along the same lines as for setup (A).

\subsection{Non-Markovian entanglement dynamics for setup (B)}
\label{sec:fourTwo}

\begin{figure}[ht]
\centering{
	\includegraphics[width=0.95\columnwidth]{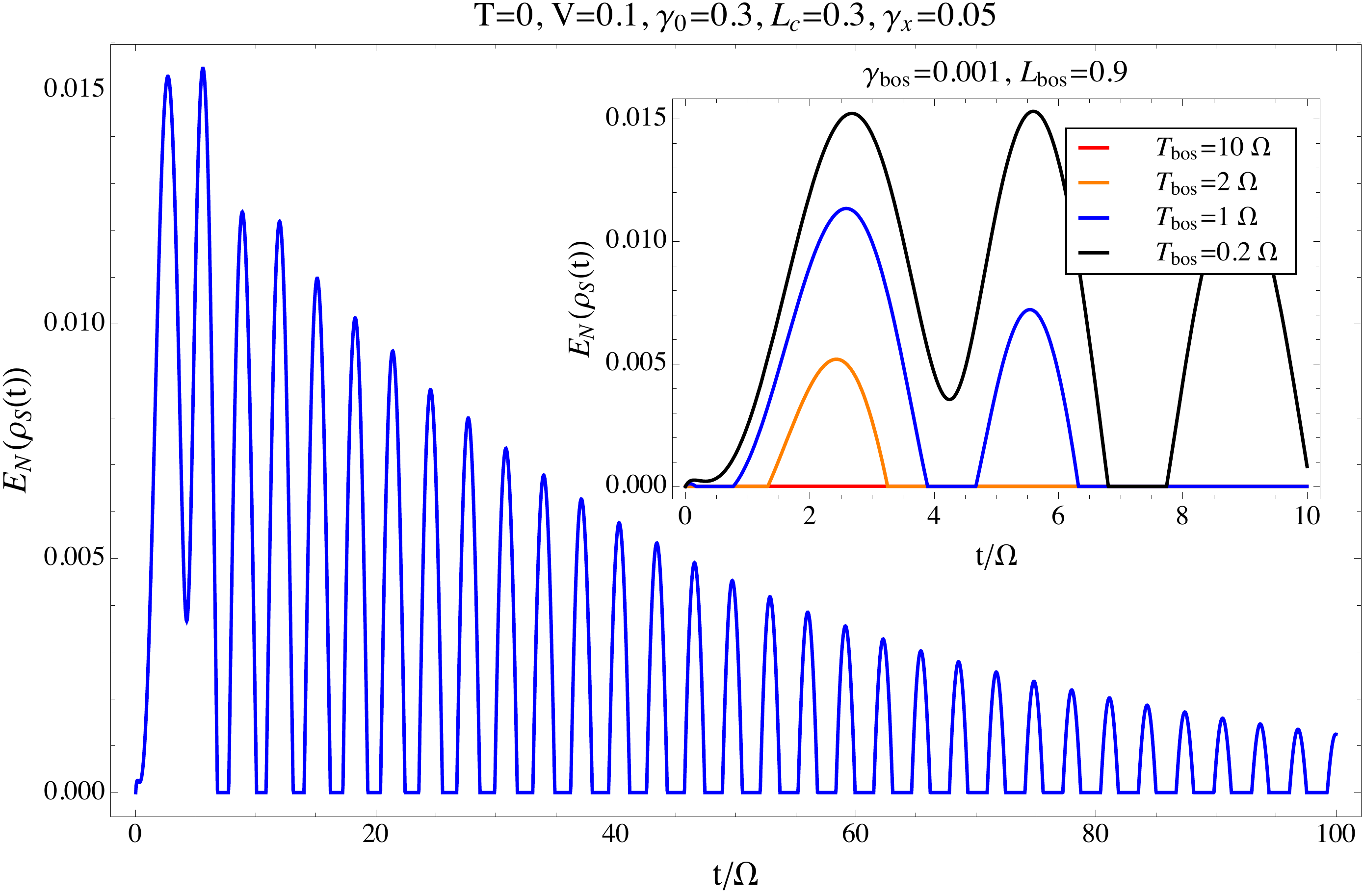}
	\caption{(Color online) Logarithmic negativity for $L_{c}=0.3$. Here, the electronic temperature is $T=0$, $V=0.1 \, \Omega$, $\gamma_{0}=0.3$,
	and $\gamma_{x}=0.05 \sqrt{m\Omega}$. The inset shows the influence of a finite bosonic heat bath with inverse temperature $\beta_{\trm{Bos}}$
	as described in the text.}
	\label{fig:SCF2}
	}
\end{figure}

In Fig.~\ref{fig:SCF2}, we present the time evolution of the degree of entanglement (for clarity depicted for
the logarithmic negativity only, but for this class of states, the entanglement of formation is a simple function of that quantity; cf. Appendix~\ref{sec:appendix})
for setup (B). We conclude that it is possible to significantly entangle two nanomechanical resonators which are coupled to three
independent reservoirs within an Andreev entangler setup, following non-Markovian open system dynamics in nonequilibrium.

Next, we discuss the influence of finite temperature on the entanglement. A non-zero temperature will render the generation
of entanglement more challenging, but the scheme is rather robust with respect to such thermal effects.  A finite electronic
temperature can be included by taking a finite value for the inverse temperature $\beta>0$ in the Fermi distribution function.
More importantly, the effects of the non-zero temperature of the substrate to which the oscillators are clamped can be included
by means of additional terms  to the master equation. This coupling can be modelled by a Caldeira-Leggett-type coupling
\begin{align}
	H_{SP_n} =  x_n\otimes \sum_{j} \gamma_{\trm{Bos,j}}  X_{n,j},
\end{align}
($n=1,2$), where each nanomechanical oscillator couples to an independent phononic heat bath (described by the position operator
$X_{n,j}$). The couplings $\{\gamma_{\trm{Bos,j}}\}$ define the spectral function
or  density in the continuum limit. This problem can be solved exactly \cite{Hu}. However, to an extraordinarily good approximation,
a very weak coupling of this form can be captured by a Markovian master equation. For low mechanical quality factors,
deviations from ohmic spectral densities and non-Markovian effects may become relevant; yet, this is not the limit we are
interested in here, but rather the limit of large mechanical quality factors and weak damping.  We can hence well
approximate the dynamics by appending the following terms to the master equation:
\begin{align}
	&\dot \rho_S(t) = {\cal L}(\rho_S(t)) \\
	&+ \gamma_{\trm{Bos}} \sum_{n=1,2} [\bar n(\beta_{\trm{Bos}}) +1] [a_n \rho_S(t)  a_n^\dagger - \frac{1}{2} \left\{ a_n^\dagger  a_n,\rho_S(t)\right\} ] \nn \\
	&+ \gamma_{\trm{Bos}} \sum_{n=1,2} \bar n(\beta_{\trm{Bos}}) 	[a_n^\dagger  \rho_S(t)  a_n  - \frac{1}{2} \left\{a_n  a_n^\dagger,\rho_S(t)\right\} ] \nn
\end{align}
for some $\gamma_{\trm{Bos}}>0$ reflecting the coupling strength, where $\bar n(\beta_{\trm{Bos}}) = (e^{\beta_{\trm{Bos}}\Omega}-1)^{-1}$
and for simplicity $\gamma_{\trm{Bos,1}} = \gamma_{\trm{Bos,2}} = \gamma_{\trm{Bos}}$. This treatment, together with the
choice of a thermal state as initial state, fairly and reasonably includes non-zero temperature to the respective setups.
Here, we introduced bosonic annihilation ${a}_{n} = (\ox_{n} \sqrt{m \Omega} + i \op_{n}/\sqrt{m\Omega} ) /\sqrt{2}$ and
creation operators ${a}_{n}^{\dag} = (\ox_{n} \sqrt{m \Omega} - i \op_{n}/\sqrt{m\Omega} ) /\sqrt{2}$.

We find that the scheme is robust with respect to the impact of finite temperatures, even though the degree of entanglement is, of course, reduced in the
presence of temperature; see the inset in Fig.~\ref{fig:SCF2}.

\subsection{Markovian limit and dissipatively generated entanglement}
\label{sec:fourFive}

So far, all entanglement generated in any of the discussed setups is entanglement due to decoherence and quantum
noise. In most instances, this quantum noise has  non-Markovian components, yet setup (B) has an interesting Markovian
regime. A direct coupling between the two oscillators is due to the second contribution of the Hamiltonian Eq.~(\ref{eqn:nemsEnt21})
(i.e.,~terms of order $\gamma_{x}^{2}$). This is a new element in the Andreev entangler setup. Interestingly, these
terms can even generate entanglement in the Markovian regime.
Any Markovian master equation is of Lindblad form
\begin{align}
	&\dot \rho_S(t) = {\cal T}(\rho_S(t)) \\
	&= i[\rho_S(t),H_S] + \sum_l [M_l \rho_S(t) M_l^\dagger -\frac{1}{2} \{ M_l^\dagger M_l,\rho_S(t) \} ] \nn \, ,
\end{align}	
with $H_S$ being Hermitian, and the $M_l$ are Lindblad operators reflecting the quantum noise. For the specific situation at
hand, the Hamiltonian is given by $H_S= \Omega ( a_1^\dagger  a_1 +  a_2^\dagger  a_2)$, whereas the Lindblad operators
vary depending on the precise context. For setup (B) in the Markovian regime, in the situation of zero electron temperature
($\beta \rightarrow \infty$) and in the absence of additional coupling to a local phononic heat bath ($\gamma_{\trm{Bos}}=0$),
there is only a single Lindblad operator, $M_1=\xi^{1/2} A$, with
\begin{align}
	\xi= \frac{\tilde\rho\pi^3 \gamma_x^4 V}{(m\Omega)^2} \, ,
\end{align}
and
\begin{align}
	A = {a}_{1}^{\dag} {a}_{2}^{\pd} + {a}_{1}^{\pd} {a}_{2}^{\dag} \, ,
\end{align}
where $\tilde{\rho} = \rho_{U} \rho_{D} \rho_{sc} \rho_{sc}$ with $\rho_{\alpha}$ and $\rho_{sc}$ being constant density of
states in lead $\alpha$ and the superconductor, respectively. We also made use that in the low bias limit, we can apply
the rotating wave approximation since the oscillators cannot be excited by the applied bias and excitations can only be
swapped between them.

If one includes in this setup a nonzero temperature and an additional coupling to an external phonon bath, in the
description in the way as explained above, there are five Lindblad operators
\begin{align}
	M_1&=\xi^{1/2} A, \\
	M_2&=\gamma_{\trm{Bos}}^{1/2} (\bar n(\beta_{\trm{Bos}})+1)^{1/2}  a_1 \, , \\
	M_3&=\gamma_{\trm{Bos}}^{1/2} (\bar n(\beta_{\trm{Bos}})+1)^{1/2}  a_2 \, ,\\
	M_2&=\gamma_{\trm{Bos}}^{1/2} \bar n(\beta_{\trm{Bos}})^{1/2}  a_1^\dagger \, , \\
	M_3&=\gamma_{\trm{Bos}}^{1/2} \bar n(\beta_{\trm{Bos}})^{1/2}  a_2^\dagger \, .
\end{align}
In any of these situations, stationary states are those states $\omega_S$ for which  ${\cal T}(\omega_S)=0$.

Stationary states of the Liouvillians -- to which the system will be driven by the dynamics and which will then be left
invariant under the noisy dynamics -- can be most easily identified by casting the Liouvillian into the standard matrix form.
Here, operators $O=\sum_{j,k} o_{j,k}|j\rangle\langle k|$ are identified with state vectors $|O\rangle = \sum_{j,k} o_{j,k}|j,k\rangle$.
Under this isomorphism, the Liouvillian takes the form
\begin{align}
	T = &-i H_S\otimes \id + i \id\otimes H_S^T \\
	&+ \sum_l \left(M_l\otimes M_l^\ast - \frac{1}{2} M_l^\dagger M_l \otimes \id- \frac{1}{2} \id \otimes M_l^T M_l^\ast \right) \, . \nn
\end{align}
Subspaces relating to stationary states relate to the kernel of $T$.

In the absence of a term in the master equation reflecting a non-zero
temperature, the kernel of $T$ is not one-dimensional, and there is no unique steady state.
Within subspaces of a fixed number of excitations, however, one encounters unique stationary states.
For example, starting in a pure state associated with the state vector
$|\psi(0)\rangle = |1,1\rangle$ (or any other state in the subspace reflecting exactly two excitations),
for long times, the system will then be driven dissipatively to an entangled steady state

\begin{align}
	\omega_S = \frac{1}{2}|1,1\rangle\langle1,1| + \frac{1}{2}
	|\psi\rangle\langle \psi|,
\end{align}
with $|\psi\rangle = (|2,0\rangle + |0,2\rangle)/\sqrt{2}$, exhibiting a degree of entanglement of $E_F(\omega_S)=1/2$ and
\begin{equation}
	E_N(\omega_S) = \log_2(3/2), 
\end{equation}
on a time scale governed by the Liouvillian gap; cf.~Appendix~\ref{sec:appendix}. Stationary states will not be unique,
however. For example, in the subspace spanned by $\{|0,1\rangle,|1,0\rangle\}$ every state that is flip symmetric under an
interchange of both modes will be a stationary state.

Unsurprisingly, this fixed point is not stable under any influence of a non-zero temperature $\beta_{\trm{Bos}}>0$. Sectors
of different particle number will then start to couple, and one can then see that the stationary states become unique (with
stationary states of full rank), but they are no longer entangled.
This observation does not imply that no entanglement is generated in a dissipative fashion, however. This setup rather
gives rise to the interesting situation in which entanglement is generated by dissipation in a nonequilibrium situation.
What is more, two effects compete: The ``good'' dissipation entangles the systems via the  effect of Cooper-pair tunneling;
in contrast, the ``bad'' dissipation related to thermal effects  rather destroys the entanglement, but on much longer time
scales. The speed of the processes is precisely  governed by the respective Liouvillian gaps. The time scale set is given
by the coupling constants $\gamma_{\trm{Bos}},\xi>0$. The damping and the mechanical quality factor are related as
$\gamma_{\trm{Bos}}=\Omega/Q$, so for  $Q=100.000$ and otherwise parameters as in Fig.\ 2 of the main text, one has
$\xi/\gamma_{\trm{Bos}}=285.67$ for $T/\Omega=0.2$. Evidently, for our purpose, it is necessary to experience a moderate
coupling to the heat bath such that $\xi > \gamma_{\trm{box}}$. In a concrete experimental setting, it might, therefore, be
necessary to actively increase $\xi$ to be able to see the nonequilibrium entanglement that we propose here. It is possible
to do so by several means, e.g., increasing the coupling strength $\gamma_x$ or varying the bias voltage $V$; cf. Eq.~(4).
To sum up, with the Andreev entangler setup we can indeed generate an entangled state even in the Markovian regime,
not only inspite of, but by means of, exploiting dissipation.

\subsection{Detection of entanglement}
One of the most challenging aspects of this (as for any other scheme of this type) is the detection of entanglement. The detection
of covariance matrix elements would be sufficient to get lower bounds for entanglement measures. One way to gain information
on these entries is to couple the two oscillators to position transducers \cite{Blencowe:2000vv} and restrict the measurement
process to only two measurements per cycle. Then, position and momentum can in principle be accessed \cite{Caves:1980aa}.
To witness entanglement, even in quantitative terms \cite{Hyllus}, however, it is sufficient to measure the experimentally more
accessible quantity \cite{GEOF} 
\begin{equation}
	\Delta(\rho_S) = \min(1,(\langle(x_1- x_2)^2\rangle + \langle(p_1+ p_2)^2\rangle)/2). 
\end{equation}
This
constitutes still a challenge, as a phase reference is necessary, but less so compared to a full reconstruction of the covariance
matrix.

\section{Summary}
\label{sec:end}
In this work, we have introduced two setups which suggest that it should be feasible to dissipatively generate entanglement
between two spatially separated nanoelectromechanical oscillators, moderated by appropriate baths. In setup (A), the two
oscillators are indirectly coupled via two common fermionic baths which must then have a rather long coherence time. With
setup (B), we introduced an entirely new way of entangling two nanomechanical oscillators in an electric setup based on
the working principle of an Andreev entangler. Here, the coherent process where a Cooper pair is split leads to an effective
coupling of the two nanomechanical oscillators. For both setups, the dissipatively generated entanglement persists over
many oscillator periods, until other dissipative processes render the state separable again. It is the hope that the present
work can contribute to the quest for entangling mechanical systems in the macroscopic domain.

\section{Acknowledgements}
We thank M.~Fuchs, L.~Krinner, T.~L.~Schmidt, and C.~Stampfer for interesting and valuable discussions and
acknowledge financial support from the DFG, Swedish Research Council, the ESF (QSpiCE), the EU (Q-Essence, SIQS, RAQUEL),
the ERC (TAQ), and the BMBF (QuOReP).

\appendix
\section{Entanglement measures}
\label{sec:appendix}

Entanglement can be reasonably quantified in terms of several natural measures of entanglement.
Specifically commonly used entanglement measures are the so-called {\it entanglement of formation} \cite{EOF}
and the (logarithmic) {\it negativity} \cite{Negativity1,Negativity1b,Negativity2,Negativity3}. Both are entanglement monotones,
which means that they satisfy the conditions a meaningful measure of entanglement should satisfy. The
logarithmic negativity is defined for states $\rho_S$ as
\begin{align}
	E_N(\rho_S) = \log_2 \|\rho_S^\Gamma\|_1,
\end{align}
where $\rho^\Gamma$ is the partial transpose of the state with respect to one part of the system. The trace-norm
$\|.\|_1$ is defined as $\|A\|_1= {\rm tr}|A|$ for operators $A$. The entanglement of formation in turn is the convex
hull of the reduced entropy function. That is to say, it is given by the infimum over all pure-state decompositions
\begin{align}
	E_F(\rho_S) =  \inf \biggl\{ \sum_j p_j E( |\psi_j\rangle\langle\psi_j| ):
	\rho_S = \sum_j p_j |\psi_j\rangle\langle\psi_j|\biggr\},
\end{align}
where $E$ is the von-Neumann entropy of the reduced state of one part.

For Gaussian bipartite states of two modes, as they are encountered in several instances of the main text, there is
a simple way to compute both of these quantities. If $\fc \in \rr^{4\times 4}$ is the covariance matrix of the state
$\rho$, the covariance matrix of $\rho^\Gamma$ is given by $\fc^\Gamma = P\sigma P$, where $P=\text{diag}(1,1,1,-1)$.
The trace norm of the partial transpose can readily be computed from the  symplectic eigenvalues $\{\tilde\lambda_-,\tilde\lambda_+\}$ of $\fc^\Gamma$, $\tilde{\lambda}_-$
being the smaller one. In this way, one arrives at the expression
\begin{align}
	E_N(\rho_S) = \max(0,- \log_2 (\tilde \lambda_-)).
\end{align}
The form obtained is specifically simple for symmetric Gaussian states with covariance matrices of the form
\begin{align}
\fc =\left(
\begin{array}{cccc}
a & 0 & c & 0\\
0 & a & 0 & -d \\
c & 0 & a & 0\\
0 & -d & 0 & a
\end{array}
\right)
\end{align}
with real entries $a,c$, and $d$, as they are encountered in particular in setup (A) of the main text. Then
\begin{align}
	\tilde\lambda_- = \left((a-c)(a-d)\right)^{1/2}.
\end{align}
At the same time, the more directly detectable so-called EPR uncertainty, for states with vanishing first moments defined as
\begin{align}
	\Delta(\rho_S) = \min\biggl(1, \frac{1}{2}(
	\langle ( x_1- x_2)^2\rangle +
	\langle ( p_1+ p_2)^2\rangle
	)\biggr),
\end{align}
is found to be the same expression,  
\begin{equation}
	\Delta(\rho_S)=\min(1, \left((a-c)(a-d)\right)^{1/2}).
\end{equation}	
This basic insight can be used
in order to tackle the challenging issue of  detection, when the EPR uncertainty can be detected by means of two
phase-sensitive joint measurements of the two oscillators. It turns out that for Gaussian states with that high degree of
symmetry, the  entanglement of formation is again given by the same quantity, up to a rescaling: One finds that \cite{GEOF}
\begin{align}
	E_F(\rho_S) = f(\Delta(\rho_S)) .
\end{align}
Here the convex and monotone decreasing function $f$ is defined as
\begin{align}
	f(x) = c_+(x) \log_2 (c_+(x)) - c_-(x) \log_2 (c_-(x)),
\end{align}
where $c_\pm(x) = (x^{-1/2} \pm x^{1/2})^2/4$. So for such symmetric Gaussian states, the EPR uncertainty is the only relevant
quantity determining both the (logarithmic) negativity and the entanglement of formation. For any of the encountered states in
setup (B), the negativity can still be computed.

\section{Entanglement dynamics}
\label{sec:appendix2}

In the idealized Markovian regime of setup (B) at zero temperature, the Liouvillian governing the dynamics will decompose into 
a direct sum of terms reflecting different excitation numbers, and the single Lindblad operator $M_1=\xi^{1/2} A$ can be written as
\begin{equation}\label{Deco}
	A=\bigoplus_{n=0}^\infty A_n.
\end{equation}
If initially a finite number of excitations is present, the time-evolved state will
only occupy a certain finite subspace as well. What is more, for large classes of initial states, the value of any entanglement monotone
can readily be given, without having to compute the actual expression. For example, for the initial state $\rho_S(0) = |1,1\rangle\langle 1,1|$
in the absence of decoherence reflecting a finite temperature, states at later times will be of the form
\begin{align}
	\rho_S (t) = &p_1(t)|1,1\rangle \langle 1,1| \\
			+ &p_2(t)(|2,0\rangle+ |0,2\rangle)(\langle 2,0|+\langle 0,2|)/2 \, , \nn
\end{align}
with suitable real time-dependent prefactors $p_1$ and $p_2$, $p_1(t)+  p_2(t)=1$. The two subspaces corresponding to the
two terms  can be locally distinguished. What is more, the state in the second term $(|2,0\rangle+ |0,2\rangle)(\langle 2,0|+\langle 0,2|)/2$
is maximally entangled. From the monotonicity property of the negativity and its convexity, one can therefore conclude that
$\|\rho_S(t)\|_1-1 = p_2(t)$ must be true, and hence
\begin{align}
	E_N(\rho_S(t)) = \log_2(\|\rho_S(t)\|_1) = \log_2(1+ p_2(t)).
\end{align}
Similarly, in this situation $E_F(\rho_S(t) ) = p_2(t)$. 

The initial preparation of $\rho_S(0) = |1,1\rangle\langle 1,1|$
 is in its own right challenging, needless to say. Still, other initial states are also conceivable, while
still arriving at entangled states. What is more, simple lower bounds for the degree entanglement can be identified from blocks of
a certain excitation number only. For example, within the section of exactly two excitations, one finds in the decomposition of Eq.\ (\ref{Deco})
the expression
\begin{eqnarray}
	A_2 
	&=&2^{1/2} \biggl( |2,0\rangle\langle 1,1| + |1,1\rangle \langle 2,0|\nonumber\\
	&+& |1,1\rangle \langle 0,2| + |0,2\rangle \langle 1,1|\biggr). 
\end{eqnarray}
Consider, e.g., the situation of one mode being prepared in a thermal state of excitation number $\bar n >0$,
the other in the ground state $|0\rangle\langle 0 |$. The expected excitation number $\bar n$ defines $\lambda\in(0,1)$ according to
\begin{equation}
	(1-\lambda)\sum_{n=0}^\infty \lambda^n n =\frac{\lambda}{1-\lambda}= \bar n.
\end{equation}
Writing the state $\rho_S(t) $ again as direct sum
\begin{equation}
	\rho_S(t) =\bigoplus_{n=0}^\infty \rho_{n}(t),
\end{equation}
one immediately finds that for the above initial condition and within the subspace of two excitations, one has
for $t\rightarrow\infty$
\begin{eqnarray}
	\rho_2 (t)&\rightarrow &
	\frac{1}{8}
	\biggl(
	3|2,0\rangle\langle 2,0| + 2 |1,1\rangle\langle1,1|+3 |0,2\rangle\langle0,2|\nonumber\\
	&-& |0,2\rangle\langle 2,0|- |2,0\rangle\langle 0,2|
	\biggr),
\end{eqnarray}
an operator that has a single eigenvalue smaller than $0$ taking the value $-1/8$. Hence,
as a simple lower bound, one finds
\begin{equation}
	E_N(\rho_S(t)) \geq \log_2\left(
	1+ \frac{1}{4}\lambda^2 (1-\lambda)\right).
\end{equation}
One can similarly also proceed for the entanglement of formation. This shows that also for initial states different from pure states, entanglement is created in the setting
when the Markovian description is largely valid.
For the non-Markovian description of setup (B) and any formulation that includes temperature,
one still obtains efficiently computable expressions for the negativity.


\bibliographystyle{apsrev}

\begin{thebibliography}{99}
\bibitem{Markovian}
	Throughout this work, we say that open systems dynamics is Markovian if it is governed by a
	time-independent Liouvillian.\ Otherwise, it is called non-Markovian \cite{NonMarkov1,NonMarkov2,NonMarkov3}.
\bibitem{setup_a}
It should be noted that -- for setup (A) -- we drop the which-path
information on the Hamiltonian level by neglecting the momentum dependence
of the tunneling amplitudes in Eq.~(\ref{eqn:nemsEnt8}). Particularly, this assumption
implies that the information through which oscillator an electron tunneled from
the left reservoir to the right one is not preserved. Therefore, on the
basis of our modeling of setup (A) we discard the finite size and the geometry 
of the setup. Effectively, the left and the right reservoirs are assumed to be
point-like. This is a rather strong assumption, particularly for the right reservoir.
\bibitem{NoteSCgap}	
	We focus here on transport at energies in the
	superconducting gap $\Delta >0$, which is the largest energy scale of the
 	problem.
\end{thebibliography}


\end{document}